\documentclass[12pt,prb,aps]{revtex4}
\usepackage[dvips]{epsfig,graphics,color}

\begin{document}
\title {Dynamical behaviors of two electrons confined in a line
shape three quantum dot molecules driven by an ac-field }
\author{ Cheng-shi Liu, Ben-kun Ma}
\affiliation{Department of Physics, Beijing Normal University,
Beijing, 100875, Peoples' Republic of China}

\begin{abstract}
Using the three-site Hubbard model and Floquet theorem, we
investigate the dynamical behaviors of two electrons confined in a
line-shape three quantum dot molecule driven by an ac electric
field. Since the Hamiltonian contains no spin-flip terms, the 6
dimension singlet state and 9 dimension triplet state sub-spaces
are decoupled and can be discussed respectively. In particular the
9 dimension triplet state sub-spaces can also be divided into 3
three-dimensional state sub-space which are fully decoupled. The
analysis shows that the Hamiltonian in each three-dimensional
triplet state sub-space as well as the singlet state sub-space in
the no double-occupancy case has the same form which is similar to
that of the driven two electron in two quantum dot molecule. By
solving the time-dependent Sch\"odinger equation, we investigate
the dynamical properties in the singlet state sub-space, and find
that the two electrons can maintain its initial localized state
when driving by an appropriately ac-field. In special, we find the
electron interaction enhances the dynamical localization effect.
Using both perturbation analytic and numerical approach to solve
the Floquet function leads to a detail understanding of this
effect.
\end{abstract}

\maketitle

PACS: 72.20.Ht; 78.66.-w; 33.80.Be

\section{Introduction}

Quantum state engineering via optical or electrical manipulation
over the coherent dynamical of suitable quantum mechanical systems
is a subject of great current interest because of a growing number
of possible experimental applications. The development of laser
and masers also open the doorway for creation of novel effects in
nonlinear quantum systems which interact with strong
electromagnetic field\cite{PR304p229}. In particular, recent
experimental successes in detecting Rabi oscillations in quantum
dot(QD) systems driven by ac-field has also spurred interest in
the use of driven ac-field to coherently manipulate the time
development of electronic states\cite{nature395p873}. Achieving an
understanding of the dynamics of the QD system is extremely
desirable, as the ability to rapidly control the localization of
electrons suggests possible application to quantum computation and
quantum information processing, in which the coherent manipulation
of entangled quantum states is essential component. An exciting
possibility is to make use the phenomenon of $coherent$
$destruction$ $of$ $tunnelling$ (CDT), in which the tunnelling
dynamics of a quantum system become suppressed at certain
parameters of the field. Tuning the driving field thus provides a
simple mechanism to localize or move charge within the quantum dot
on a rapid time-scale by destroying or restoring the tunnelling
between regions of the devices, so allowing the ac-field to be
used as "electron tweezers"\cite{cond-mat/0207317}.

In this paper, we address the dynamical localization of two
interacting electrons confined in three QD molecule with linear
arrangement driven by an ac electric field. We use an three-site
model of Hubbard-type to describe the dynamical system, which
gives a considerable computational advantage over standard
numerical approaches, and also allows us to easily include the
important effects of the electron correlations produced by the
Coulomb interaction. The two electron distribution states are
taken as basis vectors to write the Hamiltonian matrix. We show
that the system can be analyzed in singlet state and triplet state
sub-space respectively, and the triplet state sub-space can also
be divided into 3 three-dimensional syate sub-space. By solving
the time-dependent Sch\"odinger equation, we investigate the
dynamical properties in the singlet state and triplet state
subspace, and explain these finding by Floquet theorem. These
exposition are useful to understand how the ac electronic field
can affect the charge distribution inside a QD molecule, and how
they can be used for quantum control.

\section{MODEL AND METHODS}

We consider a highly simplified model in which each quantum dot is
replaced by a single site. The electron can tunnel between in the
sites, and importantly, we include interactions by means of a
Hubbard $U$-term. We do not take the energy level detuning between
the three quantum dots into account. Assuming that the AC field
frequency is much lower that single-particle level spacing, the
higher-lying single-particle states are also ignored. The
Hamiltonian defining the system read:

\begin{equation}
\begin{array}{lll}
H(t)&=&\sum\limits_{(\sigma=\uparrow\downarrow)(k=1,2,3)}\epsilon_{\sigma{k}}(t)
{d^{\dagger}_{\sigma{k}}}{d_{\sigma{k}}}+\sum\limits_{(\sigma=\uparrow\downarrow)(k=1,2)}W
({d^{\dagger}_{\sigma{k}}}{d_{\sigma{k+1}}}+h_.c_.)+\\
&&\sum\limits_{(k=1,2,3)}U_1(n_{\uparrow{k}}n_{\downarrow{k}}+
n_{\downarrow{k}}n_{\uparrow{k}})+\sum\limits_{(\sigma,{\sigma^{'}})(k=1,2)}
U_2(n_{\sigma{k}}n_{\sigma^{'}(k+1)}).
\end{array}
\label{Hamiltonian}
\end{equation}

Where $\sigma$ is the electron spin, and $\uparrow$ and
$\downarrow$ are used to indicated the up-spin and down-spin;
$k=1,2,3$ is the serial number of the QD;
${d^{\dagger}_{\sigma{k}}}$/$d_{\sigma{k}}$ are the
creation/annihilation for an electron of spin $\sigma$ on dot $k$;
The quantity $W$ denotes the hopping between adjacent dot; $U_1$
is the standard Hubbard-$U$ terms, giving the energy cost for the
double-occupation of a dot; $U_2$ presents the Coulomb repulsion
between electrons occupying neighboring dots;
$n_{\sigma{k}}=d^{\dagger}_{\sigma{k}}d_{\sigma{k}}$ is the total
charge occupation of dot $k$. For convenience, we consider the
three QD molecule is in a standing wave field, and the centra QD
is just in the wave node. We also let the QD molecule align with
propagation direction of the standing wave field. So the energy
level can be parameterized as:
$\epsilon_{\sigma{k}}(t)=\pm{(k-2)V{\cos(\omega{t})}}$, where
$\omega$ is the frequency of driving field, and $V$ is the
amplitude of energy level. Since $V$ is proportionate to the
amplitude of the ac-field, we still take $V$ as the amplitude of
the ac-field conveniently.

There are $15$ kinds of two electron distribution in the three QD
molecule which are shown in Table \ref{two electron distribution},
where $A$,$B$,$C$ indicate the site of the three QDs. We replace
the states $|5\rangle$, $|6\rangle$, $|9\rangle$, $|10\rangle$,
$|13\rangle$, $|14\rangle$ by
$|5^{'}\rangle=\frac{1}{\sqrt{2}}(|5\rangle+|6\rangle)$,
$|6^{'}\rangle=\frac{1}{\sqrt{2}}(|5\rangle-|6\rangle)$,
$|9^{'}\rangle=\frac{1}{\sqrt{2}}(|9\rangle+|10\rangle)$,
$|10^{'}\rangle=\frac{1}{\sqrt{2}}(|9\rangle-|10\rangle)$,
$|13^{'}\rangle=\frac{1}{\sqrt{2}}(|13\rangle+|14\rangle)$,
$|14^{'}\rangle=\frac{1}{\sqrt{2}}(|13\rangle-|14\rangle)$, and
take them as basis vectors. A major advantage of these is that
they manifest the two-electron spin and space distribution
clearly. We can write the Hamiltonian(Eq.\ref{Hamiltonian}) in the
space spanned by the basis vectors. Therefore the
Hamiltonian(Eq.\ref{Hamiltonian}) can be described by $15\times15$
matrix. The Hamiltonian(Eq.\ref{Hamiltonian}) contains no
spin-flip terms since measurements on semiconductor QD show that
the spin-flip relaxation time is typically extremely long, and so
the singlet state and triplet state sub-space are completely
decoupled. Thus we can investigate the dynamical behaviors of two
electrons in two sub-space respectively. Therefore, if the initial
state posses a definite party this will be retained throughout its
time evolution, and we only need to include states of the same
party in the basis.

The Hamiltonian(Eq.\ref{Hamiltonian}) also contains no
spin-orbital coupled terms, and so the wave function can be
written as the direct product of spin and orbital wave function. A
many-particle basis can then be constructed by taking Slater
determinants of single particle states defined on the QD. The
basis vectors ($|4\rangle$, $|12\rangle$, $|8\rangle$) constitute
the up-spin sub-space, and the basis vectors ($|7\rangle$,
$|15\rangle$, $|11\rangle$) constitute the down-spin sub-space.
Their spin wave function are symmetric under particle exchange.
The orbital wave function of basis vector ($|6^{'}\rangle$,
$|14^{'}\rangle$, $|10^{'}\rangle$) are antisymmetric under
particle exchange, so their spin wave function are symmetric.
Therefore, the basis vectors ($|4\rangle$, $|12\rangle$,
$|8\rangle$, $|7\rangle$, $|15\rangle$, $|11\rangle$,
$|6^{'}\rangle$, $|14^{'}\rangle$, $|10^{'}\rangle$) constitute
9-dimension spin-triplet state sub-space. The calculation also
shows that the 3 three-dimensional spin symmetric sub-space are
completely decoupled, and their Hamiltonian spanned by three
spin-symmetric basis vectors hold the same form which is

\begin{equation}
H(t)=\left(
\begin{array}{cccccc}
U_2-V\cos(\omega{t})&W&0\\
W&0&W\\
0&W&U_2+V\cos(\omega{t})
\end{array}
\right).
\label{Hamilton matrix triplet}
\end{equation}

So the dynamical behaviors of the two electrons are identical in
the 3 three-dimensional sub-space. For simple, we only investigate
the dynamical characters of two electrons in up-spin sub-space
($|4\rangle$, $|12\rangle$, $|8\rangle$).

In order to obtain the dynamical characters of the three-level
system, it is necessary to examine the system of two electrons
confined by two QD molecule driven by ac-field, which has been
fully investigated in
Ref.\cite{PLA271p491,JPCM12p2351,PRB65p113304,PRA66p022117}. In
this system, the basis vectors ($|\bar{1},\bar{1}\rangle$,
$(|\bar{1},1\rangle-|1,\bar{1}\rangle)/\sqrt{2}$, $|1,1\rangle$)
constitute the triplet stste sub-space, and the basis vectors
($|2,0\rangle$, $(|\bar{1},1\rangle+|1,\bar{1}\rangle)/\sqrt{2}$,
$|0,2\rangle$) constitute the singlet state sub-space. The two
sub-space are also decoupled. In the triplet stste sub-space, the
two-particle basis vectors ($|\bar{1},\bar{1}\rangle$,
$(|\bar{1},1\rangle-|1,\bar{1}\rangle)/\sqrt{2}$, $|1,1\rangle$)
are the eigenvectors of its Hamiltonian and constitute the trivial
triplet state sub-space in which the electron number on each QD is
invariably one, and the time-dependent term does not influence
this characteristics. In the singlet state sub-space, the
Hamiltonian is similar to that of Formula (Eq.\ref{Hamilton matrix
triplet}). The relationships of the parameters and the basis
vectors between the two system are: $U_2\rightarrow{U_1-U_2}$,
$W\rightarrow{\sqrt{2}W}$, $|4\rangle\rightarrow|2,0\rangle$,
$|8\rangle\rightarrow|0,2\rangle$,
$|12\rangle\rightarrow(|\bar{1},1\rangle+|1,\bar{1}\rangle)/\sqrt{2}$.

\section{RESULTS}

\subsection{The two-electron dynamical properties in the triplet state subspace}

We now investigate the dynamical properties of the
three-dimensional triplet subspace and take the subspace
($|4\rangle$, $|12\rangle$, $|8\rangle$) for example. The time
periodicity of Hamiltonian(Eq.\ref{Hamilton matrix triplet})
enables us to describes the dynamics within the Floquet formalism.
In addition, since the Hamiltonian is invariant under the combined
dynamic parity operation $z\rightarrow-z$;
$t\rightarrow{t+\pi/\omega}$, each Floquet state is either odd or
even. Quasi-energies of different party may cross, otherwise an
avoided crossing may occurs\cite{PR304p229}. We calculate the
quasi-energies and $P_{min}$ as function of $V$ with the
parameters $U_2=n\omega$. We used two method to obtain
quasi-energies: (1) the numerical method\cite{PR304p229} which is
to diagonalize the unitary time-evolution operator for one period
of the driving field $U(t+T,t)$. It may be easily shown that the
eigenvectors of this operator are equal to the Floquet states, and
its eigenvalues are related to the quasi-energies via
$\lambda_j=\exp(-i\varepsilon_jT)$. This method is particularly
well-suited to our approach, as $U(T,0)$ can be obtained by
integrating the unit matrix in time over one period of the field
using the Runge-Kutta method. (2) the perturbation
approach\cite{PRB65p113304,ZPB59p251}. In this method, the
quasi-energies are obtained by first solving the Floquet
equation\cite{PR304p229} in the absence of tunnelling terms, and
then performing perturbation theory with the tunnelling terms as
the "perturbation". The perturbation solution are:
$\varepsilon_{1,2}=\pm\sqrt{2}J_n(V/\omega)W$, $\varepsilon_3=0$.
where, $J_n$ is the bessel function of the $n$ order. We now
define the probability $P_{min}$. In this singlet state sub-space,
the wave function can be written as a superposition of basis
vectors:
$|\psi(t)\rangle=\sum\limits_{\alpha}C_{\alpha}(t)|\alpha\rangle$,
$|\alpha\rangle$ is the basis vector and $\alpha=4,12,8$.
Substituting this expansion into the time-dependent Schr\"odinger
equation yields a first-order differential equation for the
expansion coefficients $C_{\alpha}$ $(\hbar=1)$:

\begin{equation}
i\frac{\partial}{\partial t} \left(
\begin{array}{ccc}
C_4\\
C_{12}\\
C_4
\end{array}
\right)= H(t)\left(
\begin{array}{ccc}
C_4\\
C_{12}\\
C_4
\end{array}
\right). \label{differential equation_triplet}
\end{equation}

When the initial condition $C_\alpha(0)$ are given, we use the
Runge-kutta method to solve the differential equation
(Eq.\ref{differential equation_triplet}) and obtain the expansion
coefficients $C_{\alpha}(t)$. When $C_4(0)=1$, $C_{12}(0)=0$,
$C_{8}(0)=0$, We term the minimum value of $|C_4(t)|^2$ of
evolving $25$ driving period $LP_{min}$. When $C_4(0)=0$,
$C_{12}(0)=1$, $C_{8}(0)=0$, We term the minimum value of
$|C_{12}(t)|^2$ of evolving $25$ driving period $CP_{min}$.
$LP_{min}$ and $CP_{min}$ are used to quantify the degree of
quantum tunnelling. If $LP_{min}=1$ and $CP_{min}=1$, it means
that the two electrons can maintain its initial localization state
in a short time span, and the tunnelling between different
electron distribution is suppressed completely. If $P_{min}=0$ (or
$CP_{min}=0$), it means that the two electrons can not maintain
its initial state.

\begin{figure}[htbp]
\begin{center}
\resizebox*{5.2992in}{2.3622in}{\includegraphics{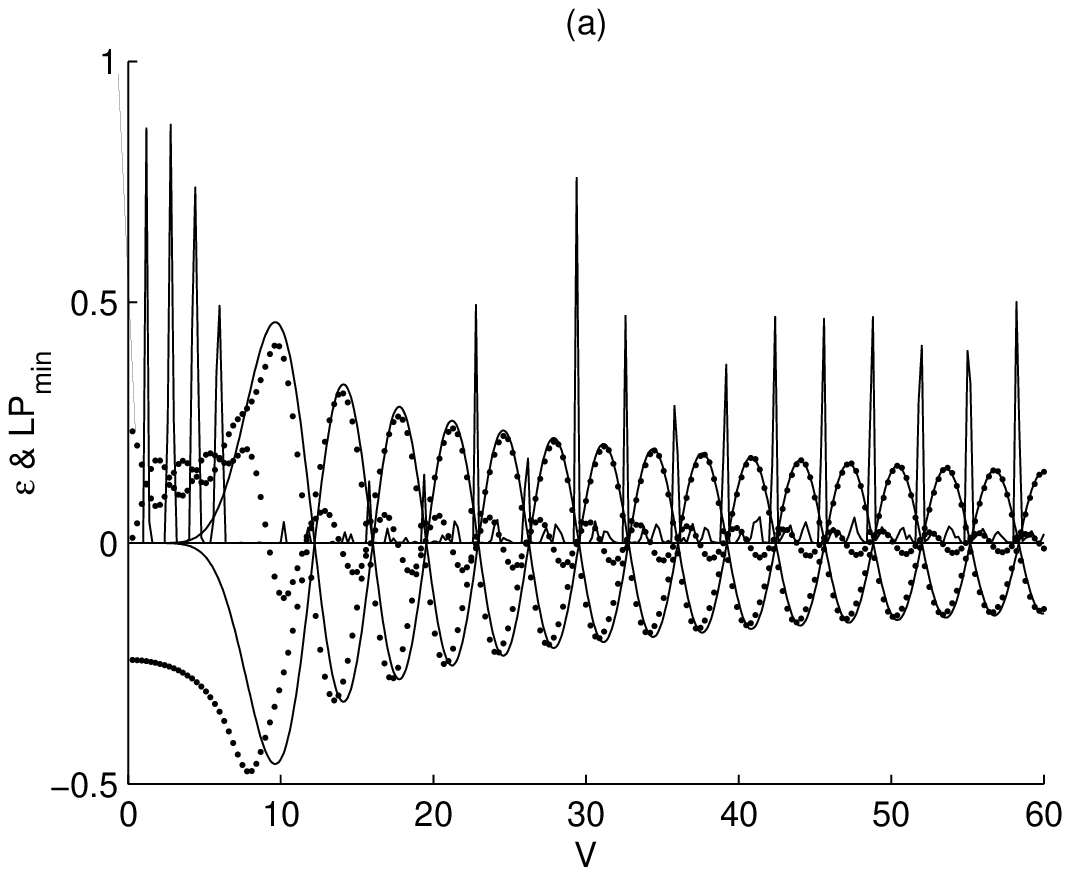}
\includegraphics{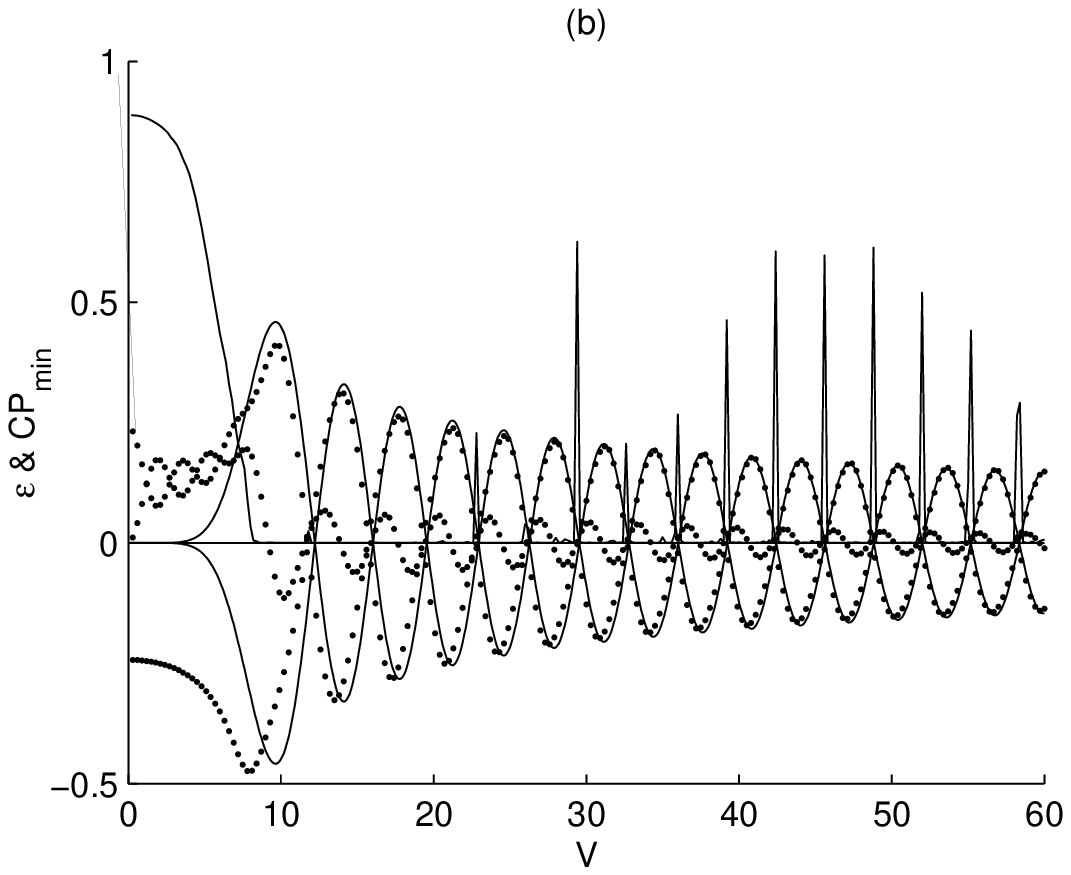}}
\caption{The dependence of quasi-energies and (a) $LP_{min}$, (b)
$CP_{min}$ as function of $V$, with the parameters: $U_2=8$,
$W=1$, $\omega=1$. dotted line = quasi-energy exact results,
dotted line = quasi-energy exact results, Solid line =
perturbation theory.} \label{3_epsilon_LPmin_CPmin_V}
\end{center}
\end{figure}

We calculate the quasi-energies $\varepsilon$ and the quantity
$LP_{min}$ and $CP_{min}$ in Fig.\ref{3_epsilon_LPmin_CPmin_V} as
a function of $V$ with the parameters $U_2=8$, $W=1$, $\omega=1$.
The calculation show that the excellent agreement between the
quasi-energies exact solution and perturbation solution for strong
and moderate fields. For weak fields, however, the driving terms
do not dominate the tunnelling terms and the perturbation theory
breaks down. Figure \ref{3_epsilon_LPmin_CPmin_V} also shows that
the dynamical properties can be divided into two regime: the
strong field regime and the weak field regime. In the strong field
regime, the peaks in $LP_{min}$ and $CP_{min}$ is found by
locating the roots of $J_8(V/\omega)$. So the initial states
$|4\rangle$, $|12\rangle$ and $|8\rangle$ can be maintained. Hence
the tunnelling between different electric distributions is
depressed. In the weak field, $CP_{min}$ decays smoothly to zero,
and a series of peak of $LP_{min}$ is found at the quasi-energy
crossing. So the initial state $|12\rangle$ fail to hold on when
increasing the driving ac field. However the initial state
$|4\rangle$ and $|8\rangle$ keep its initial state even though the
strong interaction between them.

To interpret the dynamical behaviors for the strong field regime,
we seek the perturbation approach\cite{PRB65p113304}. We treat the
tunnelling terms in Hamiltonian (Eq.\ref{Hamilton matrix triplet})
as a perturbation. Solving the Floquet equation, the degeneracy
Floquet states of the zeroth-order approximation are given by
($U_2=n\omega$):

\begin{equation}
\left\{
\begin{array}{lll}
|u_1(t)\rangle&=&|\exp[-iU_2t+i\frac{V}{\omega}\sin{\omega{t}}],0,0\rangle\\
|u_2(t)\rangle&=&|0,1,0\rangle\\
|u_3(t)\rangle&=&|0,0,\exp[-iU_2t+i\frac{V}{\omega}\sin{\omega{t}}]\rangle
\end{array}
\right.
\label{Floquet states}
\end{equation}

With the degeneracy Floquet states, we obtained the perturbing
operator $P_{ij}=\langle\langle{u_i(t)}|H_t|u_j(t)\rangle\rangle$,
where $\langle\langle{\cdot\cdot\cdot}\rangle\rangle$ denotes the
inner product in the extended Hilbert space\cite{ZPB59p251}. By
using the identity: $\exp[-i\beta\sin\omega{t}]=
\sum\limits_{m=-\infty}^{\infty}J_m(\beta)\exp[-im\omega{t}]$ to
rewrite the form of $|u_i(t)\rangle$, the matrix element of $P$
can be found to be $P_{12}=P_{23}=WJ_n(\frac{V}{\omega})$. When
the ratio of the field strength to the frequency is a root of
Bessel function $J_n$, the transition between two Floquet states
$|4\rangle$ and $|12\rangle$ as well as $|12\rangle$ and
$|8\rangle$ is forbidden. Therefore, the tunnelling between
different electrical distribution is depressed intensively, and
the phenomenon of CDT occurs.

\begin{figure}[htbp]
\begin{center}
\resizebox*{5.2992in}{2.3622in}{\includegraphics{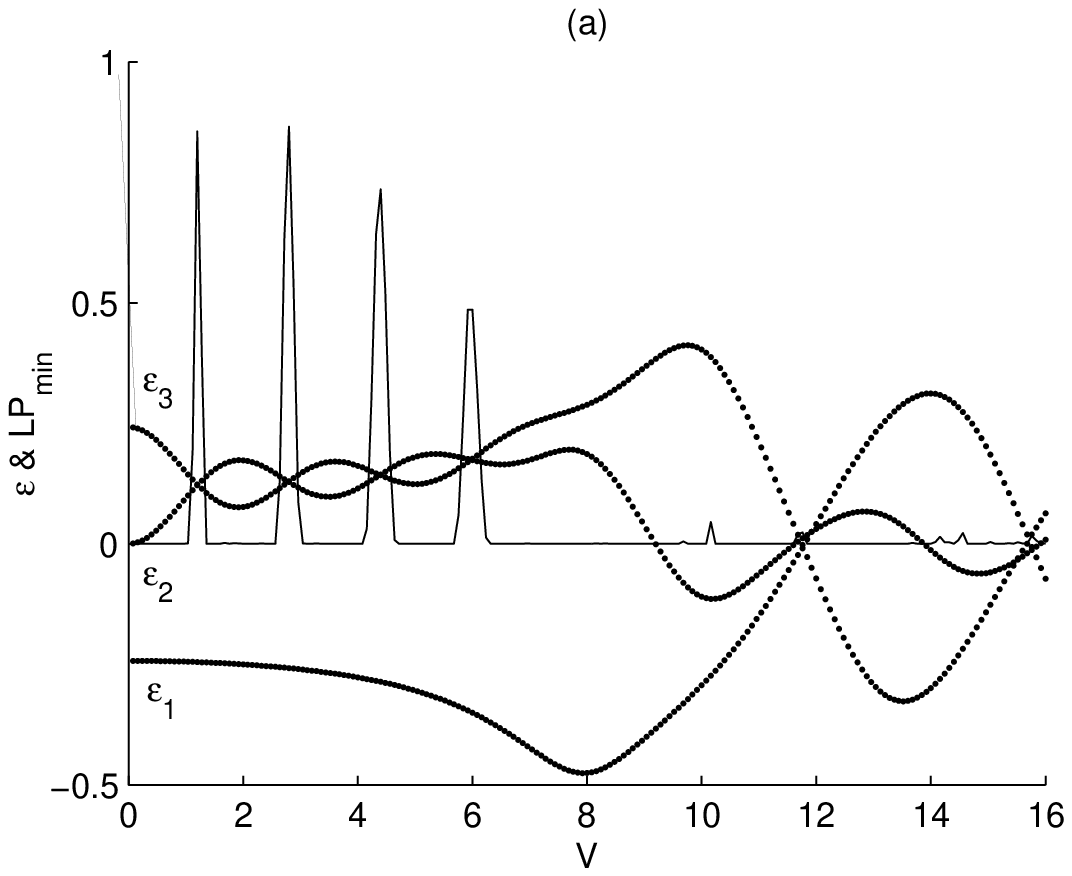}
\includegraphics{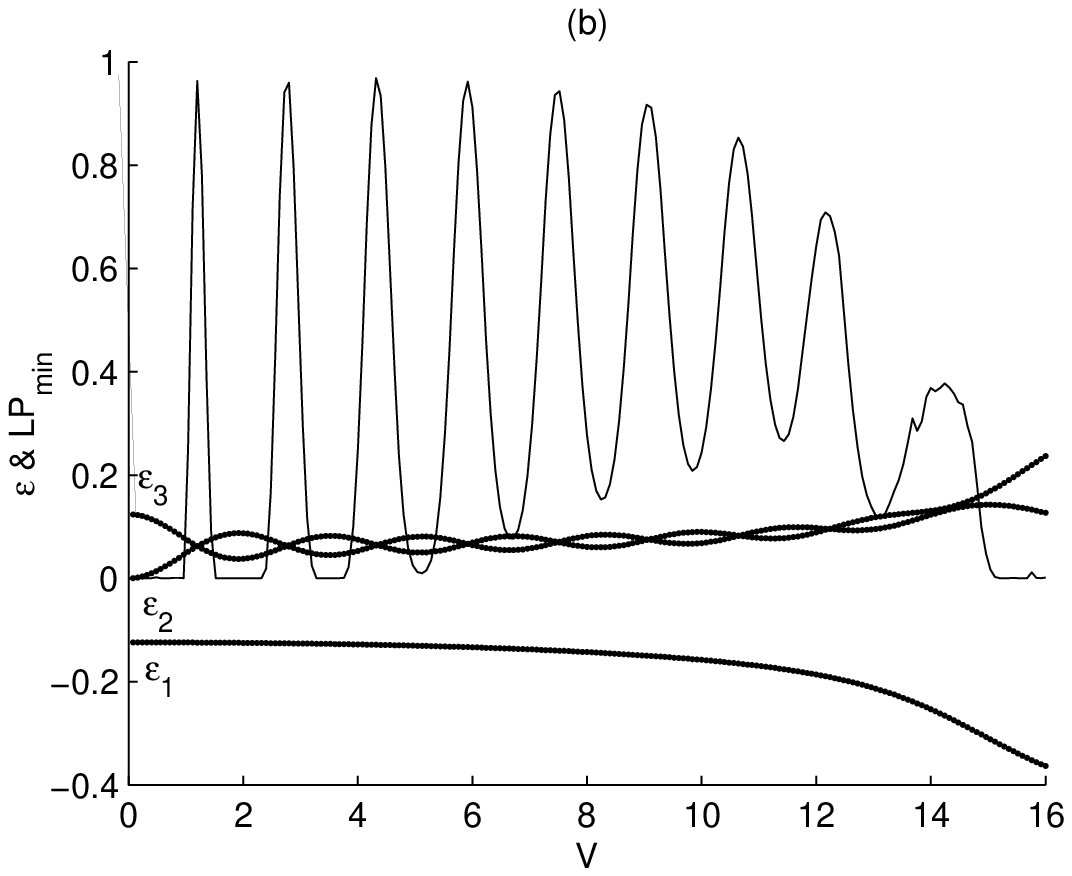}}
\caption{(a) magnified view of the weak field regime in
Fig.\ref{3_epsilon_LPmin_CPmin_V}(a), (b) the quasi-energies and
$LP_{min}$ as function of $V$, with the parameters: $U_2=16$,
$W=1$, $\omega=1$. dotted line = quasi-energy exact results,
dotted line = quasi-energy exact results, Solid line =
perturbation theory.} \label{W_epsilon_LPmin_V}
\end{center}
\end{figure}

Let us analyze the dynamical properties in the weak field regime.
In order to show it clearly, we show in
Fig.\ref{W_epsilon_LPmin_V}(a) the magnified view of
Fig.\ref{3_epsilon_LPmin_CPmin_V}(a) for the weak field regime.
The Fig.\ref{W_epsilon_LPmin_V}(a) shows that the dynamical
localization occurs at the crossing of quasi-energies
$\varepsilon_2$ and  $\varepsilon_2$. In the absent of an external
driving field, the eigenvalues and eigenvectors are

\begin{equation}
\left\{\begin{array}{ll}
|\varphi_{1}\rangle=|1,-a/(\sqrt{2}W),1>,\hspace{0.5cm}&E_1=
b=(-\sqrt{U_2^2+16W^2}+U_2)/2\\
|\varphi_{2}\rangle=|-1,0,1>,\hspace{0.5cm}&E_2=U_2\\
|\varphi_{3}\rangle=|1,-b/(\sqrt{2}W),1>,\hspace{0.5cm}&E_3=
a=(\sqrt{U_2^2+16W^2}+U_2)/2
\end{array}
\right. \label{eigenvalue and eigenvector}
\end{equation}

When driving by weak field, the two-electron Floquet states are
approximated by their eigen-states. We note $U_2\gg{W}$, and the
eigen-state $|\varphi_{3}>$ has small component of two-electron
state $|12\rangle$, so the eigen-state $|\varphi_{2}>$ and
$|\varphi_{3}>$ are similar to the excited state (asymmetry) and
the ground state (symmetry) of a electron in two QD. Thus it
expects the initial two-electron state $|4\rangle$ and $|8\rangle$
remains forever at the crossing of quasi-energies $\varepsilon_2$
and $\varepsilon_3$\cite{PLA271p491,JPCM12p2351,PRA66p022117}. The
rate $V/\omega$ at the first crossing is about $1.2$, a root of
the zero-order Bessel function, suggesting that this kind of
dynamical properties can be approximates by a two-level model.
When increasing the interaction $U_2$, the component of state
$|12\rangle$ in $|\varphi_{3}\rangle$ will decrease. So the
Coulomb repulsion may help to maintain the initial states
$|4\rangle$ and $|8\rangle$. We present the calculation in
Fig.\ref{W_epsilon_LPmin_V}(b) with $U_2=16$, and the other
parameters is the same as that of Fig.\ref{W_epsilon_LPmin_V}(a).
The calculation in Fig.\ref{W_epsilon_LPmin_V}(b) shows that a
series peaks of $LP_{min}$ emerge, and its height increase
obviously with the increasing of $U_2$.

\subsection{The two-electron dynamical properties in the singlet state subspace}

We now investigate the dynamical behaviors for the singlet state
sub-space. The orbital wave function of the basis vectors
($|1\rangle$, $|2\rangle$, $|3\rangle$) are symmetric under the
electron exchange, so their spin wave function are antisymmetric.
The spin wave function of the basis vectors ($|9^{'}\rangle$,
$|5^{'}\rangle$, $|13^{'}\rangle$) are antisymmetric under the
electron exchange, so their orbital wave function are symmetric.
Therefore, the basis vectors ($|9^{'}\rangle$, $|2\rangle$,
$|5^{'}\rangle$, $|3\rangle$, $|13^{'}\rangle$, $|1\rangle$)
constitute the singlet sub-space. The
Hamiltonian(\ref{Hamiltonian}) spanned by the six
spin-antisymmetric basis vectors can be written as $6\times6$
matrix, which is

\begin{equation}H(t)=
\left(
\begin{array}{cccccc}
U_2+V\cos(\omega{t})&\sqrt{2}W&0&\sqrt{2}W&W&0\\
\sqrt{2}W&U_1&\sqrt{2}W&0&0&0\\
0&\sqrt{2}W&U_2+V\cos(\omega{t})&0&W&\sqrt{2}W\\
\sqrt{2}W&0&0&U_1+2V\cos(\omega{t})&0&0\\
W&0&W&0&0&0\\
0&0&\sqrt{2}W&0&0&U_1-2V\cos(\omega{t})
\end{array}
\right).\label{Hamilton matrix singlet}
\end{equation}

Due to the diversity of the size and couple between the QDs, the
dynamical behaviors of the system show multiplicity. We begin our
investigation by first considering the simplest case, that of the
Hubbard $U_1$-term to be infinitely large - that is, we work in
the sub-space of states with no double occupation. Our Hilbert
space is thus three-dimensional spanned by the basis vectors
($|5^{'}\rangle$, $|13^{'}\rangle$, $|9^{'}\rangle$). Therefore,
the Hamiltonian can be written as:

\begin{equation}
H(t)=\left(
\begin{array}{cccccc}
U_2-V\cos(\omega{t})&W&0\\
W&0&W\\
0&W&U_2+V\cos(\omega{t})
\end{array}
\right) \label{Hamilton matrix interaction no double-occupancy}
\end{equation}

This form of Hamiltonian(Eq.\ref{Hamilton matrix interaction no
double-occupancy}) is the same as that of
Hamiltonian(Eq.\ref{Hamilton matrix triplet}). Thus the six-level
system can be simplified to the three-level system, and the
dynamical characters in this special case is also similar to two
electrons in two QD molecule.

We now take the most general case, and consider the $U_1$ and
$U_2$ to be a finite value which means that the three
double-occupation are no longer energetically exclude from the
dynamics, and accordingly we must take the full six-dimensional
basis set. we calculate the quasi-energies and $P_{min}$ as
function of $V$ with the parameters $U_1=n_1\omega$ and
$U_1=n_2\omega$. We used two method to obtain quasi-energies: (1)
the numerical method; (2) the perturbation approach, which are
similar to the calculation of Figure
\ref{3_epsilon_LPmin_CPmin_V}. The perturbation solution of
quasi-energies are

\begin{equation}\left\{
\begin{array}{lll}
\varepsilon_{1,2}&=&\pm\sqrt{2}J_{n_1-n_2}W\\
\varepsilon_{3,4}&=&\pm\sqrt{2J_{n_2}^2+6J_{n_1-n_2}^2}W\\
\varepsilon_{5}&=&0
\end{array}
\right.
\end{equation}

where, $J_{n_1-n_2}$ and $J_{n_2}$ are the bessel function of the
$n_1-n_2$ and $n_2$ order respectively.

We now define the probability $P_{min}$ in the singlet sub-space.
The wave function can be written as a superposition of basis
vectors:
$|\psi(t)\rangle=\sum\limits_{\alpha}C_{\alpha}(t)|\alpha\rangle$,
$|\alpha\rangle$ is the basis vector and
$\alpha=9^{'},\cdot\cdot\cdot,1$. Substituting this expansion into
the time-dependent Schr\"odinger equation yields a first-order
differential equation for the expansion coefficients $C_{\alpha}$
$(\hbar=1)$:

\begin{equation}
i\frac{\partial}{\partial t} \left(
\begin{array}{ccc}
C_{9^{'}}\\
\cdot\\
\cdot\\
\cdot\\
C_1
\end{array}
\right)= H(t)\left(
\begin{array}{ccc}
C_{9^{'}}\\
\cdot\\
\cdot\\
\cdot\\
C_1
\end{array}
\right). \label{differential equation_singlet}
\end{equation}

When the initial condition $C_\alpha(0)$ are given, we solve the
differential equation(Eq.\ref{differential equation_singlet}) by
using the Runge-kutta method to obtain the expansion coefficients
$C_{\alpha}(t)$. We term the minimum value of $|C_{\alpha}(t)|^2$
of evolving $25$ driving period $P_{min}$, and use it to quantify
the degree to which the ac-field bring about localization. In
special, we term the minimum value of $|C_1(t)|^2$ of evolving
$25$ driving period $AP_{min}$ when $C_1(0)=1$, $C_{\alpha}(0)=0$
$\alpha\neq1$. We also term the minimum value of $|C_2(t)|^2$ of
evolving $25$ driving period $BP_{min}$ with the initial condition
$C_2(0)=1$, $C_{\alpha}(0)=0$ $\alpha\neq2$. If $AP_{min}=1$, it
means that the two electrons can maintain its initial localization
state (the left QD A) in a short time span, and the tunnelling
between different electron distribution is suppressed completely,
so the phenomenon occurs. If $AP_{min}=0$, it means that the two
electrons can not maintain its initial localization state (the
left QD).

\begin{figure}[htbp]
\begin{center}
\resizebox*{5.2992in}{2.3622in}{\includegraphics{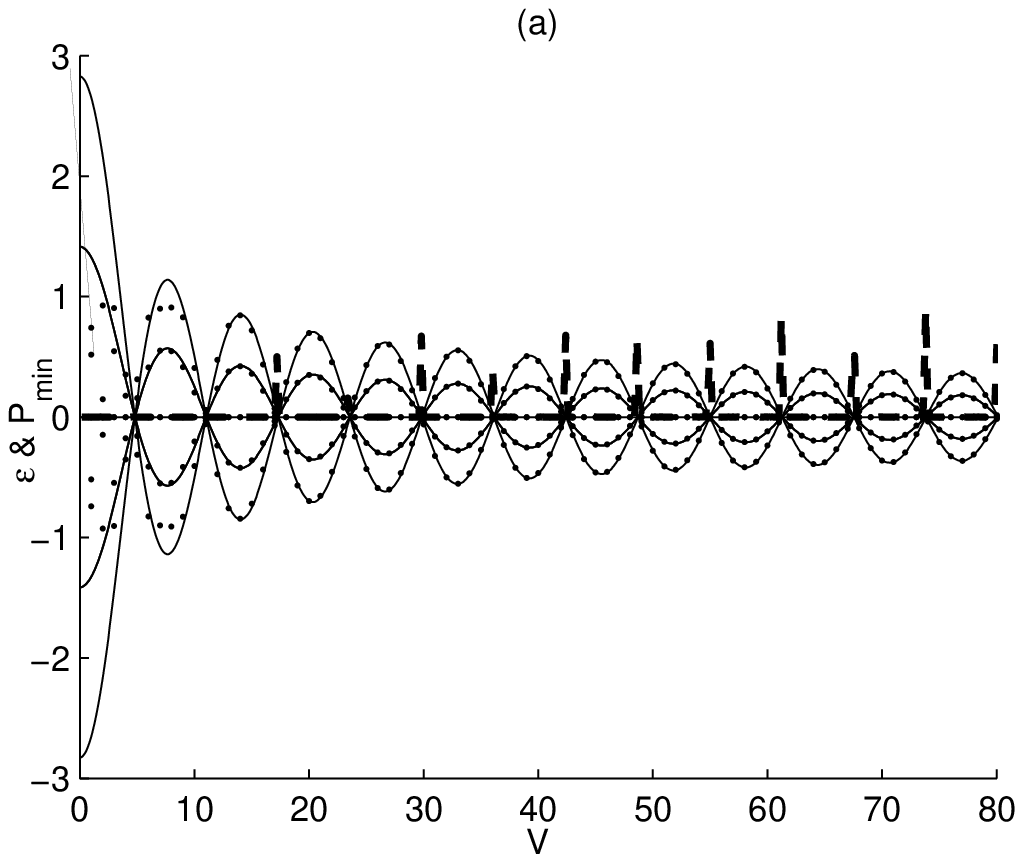}
\includegraphics{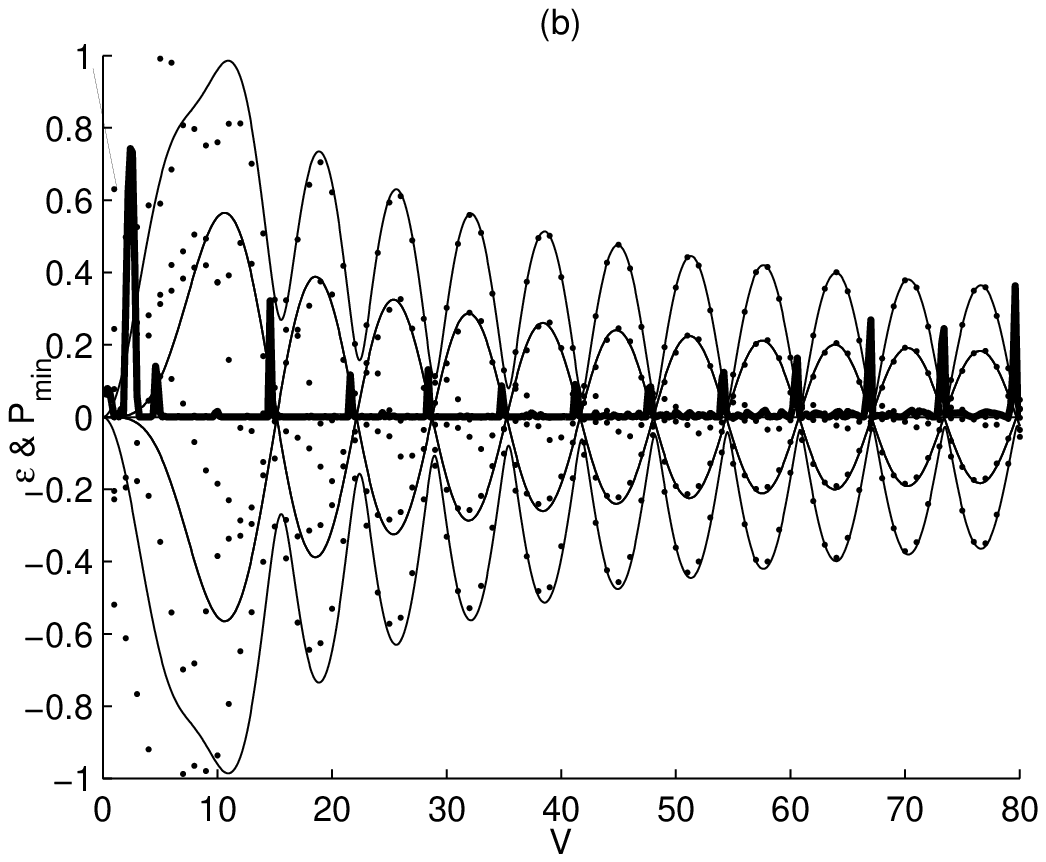}}

\resizebox*{5.2992in}{2.3622in}{\includegraphics{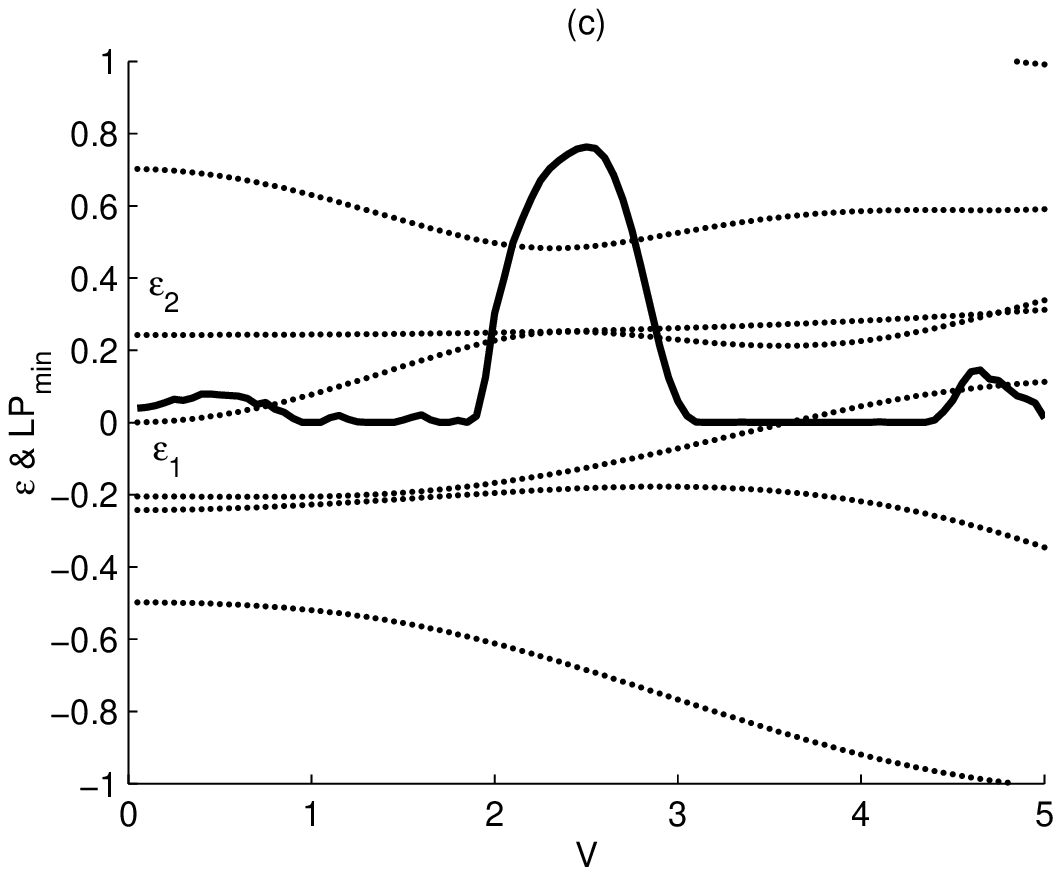}
\includegraphics{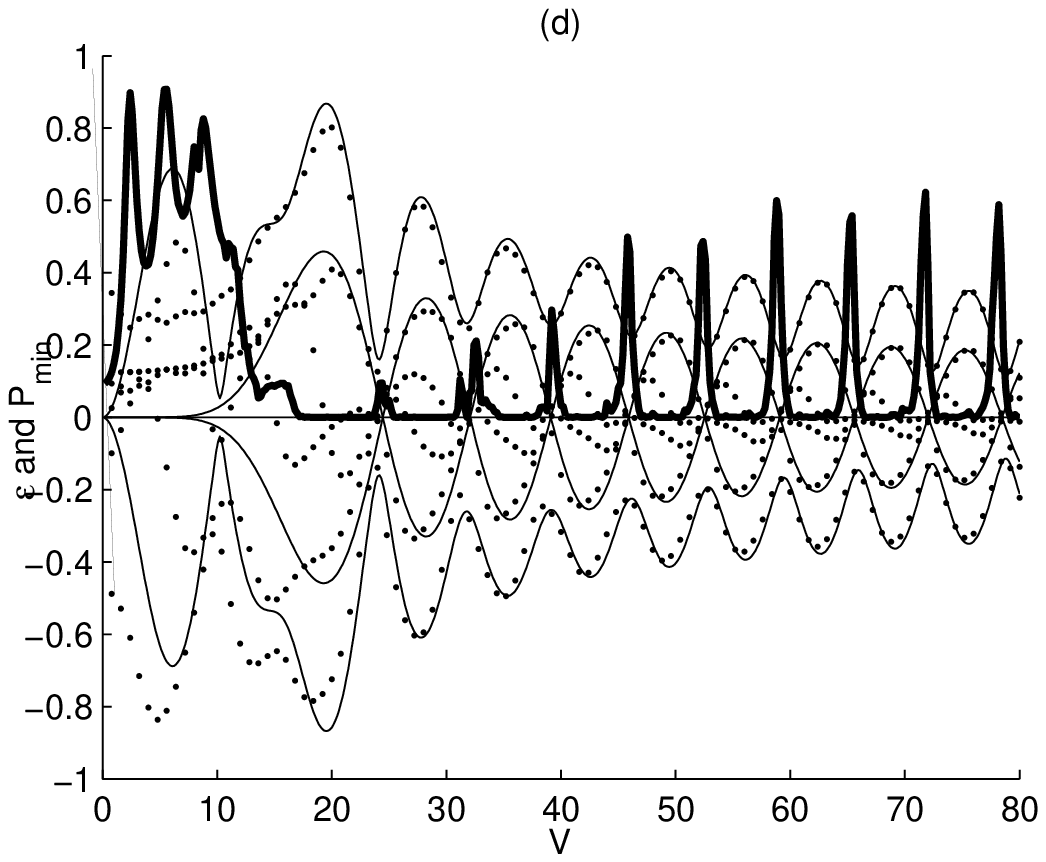}}

\caption{The dependence of $AP_{min}$ as function of $V$ with the
parameters: (a) $U_1=0$, $U_2=0$, $W=1$, $\omega=2$;  (b)
$U_1=12$, $U_2=4$, $W=1$, $\omega=2$. (c) magnified view of
Fig.\ref{epsilon_LPmin_V}(b) for the weak field regime. (d) The
dependence of $BP_{min}$ as function of $V$ with the parameters
$U_1=20$, $U_2=4$, $W=1$, $\omega=2$.} \label{epsilon_LPmin_V}
\end{center}
\end{figure}

In order to show the dynamical properties in this general case, it
is necessary to show the case of non-interacting and
double-occupancy permitted firstly, and then investigate how the
Coulomb interaction affects its dynamical behaviors. The three
sites model in this non-interaction case is similar to that
non-interaction electrons driven by ac-field in a
superlattice\cite{ZPB59p251,PBL69p351}, with the quasi-energy
crossing corresponding to "miniband collapse". We term the
phenomenon as "quasi-energy collapse". The electron tunnelling
between QD is considerably suppressed with parameters at the roots
of $J_0(V/\omega)$ ($J_0$ is the Bessel function of first kind).
The calculation in Figure \ref{epsilon_LPmin_V}(a) verifies our
assertion. We show in Figure \ref{epsilon_LPmin_V}(a) the
quasi-energy spectrum and $AP_{min}$ obtained by sweeping over
$V$. The dot and slimline present the exact solution and the
perturbation solution of quasi-energies respectively, and the
thick line presents $AP_{min}$. Figure \ref{epsilon_LPmin_V}(a)
demonstrates the excellent agreement between the exact solution
and perturbation solution of the system quasi-energies for strong
and moderate fields. For weak fields, however, the driving terms
do not dominate the tunnelling terms and the perturbation theory
breaks down. Figure \ref{epsilon_LPmin_V}(a) also shows that the
peaks in $AP_{min}$ by locating at the point of quasi-energies
crossing. Hence the phenomenon of CDT occur at the point of
quasi-crossing.

When the electron interactions involving, the difference from the
non-interaction case arise. Using the same method as that of
Figure \ref{epsilon_LPmin_V}(a), we calculate in Figure
\ref{epsilon_LPmin_V}(b) the quasi-energy and $AP_{min}$ as a
function of $V$ with $U_1=12$ and $U_2=4$. It shows that the
phenomena which is similar to "miniband collapse" occurs in the
strong-field regime of the quasi-energy, even though not all the
quasi-energy levels cross at one point. At the weak-field regime,
the feature of quasi-energy (dotted line) is different obviously
from that of the non-interaction case, and a series of
quasi-energy crossing appears. The calculation in Figure
\ref{epsilon_LPmin_V}(b) also shows that $AP_{min}-V$ can be
divided apparently into two regime: the weak-field regime and the
strong-field regime. At the strong-field regime, the character of
$AP_{min}-V$ is similar to that of the non-interaction case
(Fig.\ref{epsilon_LPmin_V}(a)). At the weak field regime, it is
interesting to see that some sharp peaks of $P_{min}$ occur at the
quasi-energy crossing. The results indicate that two electrons
localized in left QD initially can maintain its localization state
(a $25$ driving period) when driving by a weak-field, even though
there is a strong Coulomb interaction. Therefore, the Coulomb
interactions have little effect on strong-field regime, and they
enhance the phenomena of CDT in the weak-field regime. This
quantum phenomena can also occur in two electrons in two QD
molecule, and have been found and analyzed in Ref
\cite{PRA66p022117}.

We now analyze the system dynamical properties when driving by
weak field. In order to show it clearly, we give in
Fig.\ref{epsilon_LPmin_V}(c) the magnified view of
Fig.\ref{epsilon_LPmin_V}(b) for the weak field regime.
Fig.\ref{epsilon_LPmin_V}(b) demonstrates that the position of the
peak in $LP_{min}$ to be found by locating at the crossing of
quasi-energies $\varepsilon_1$ and $\varepsilon_2$. When driving
by weak field, the Floquet states can be approximated by the
corresponding unperturbed eigen-states. We diagonalize the
Hamiltonian (Eq.\ref{Hamilton matrix singlet}) when $V=0$, and
obtain $\varphi_1=(-0.5774,0.5774,-0.5774,0,0,0)$ and
$\varphi_2=(-0.6969,0,0.6969,0.1196,-0.1196,0)\approx(-0.6969,0,0.6969,0,0,0)$.
The two eigen-states are similar to the symmetry (ground state)
and asymmetry (the first excited state) states of a single
electron in three QD. This is also like the case in the driven
two-level model. Moreover, the value of the parameters
$2V/\omega=2.4$, which is the root of the zero-order Bessel
function, suggesting that in this situation can be approximated by
the driven two-level model.

We apply the perturbation
approach\cite{cond-mat/0207317,PRB65p113304} to analyze the
dynamical property for the strong field regime. When driven by
strong field, the Hamiltonian(Eq.\ref{Hamilton matrix singlet})
can be divided into two parts: $H_t$ which contain all the
tunnelling terms, and $H_I$ containing all interaction terms
(these involving $U_1$ and $U_2$ and the electric field). We then
find the eigen-system of the operator
$H_I(t)=H_I-i(\partial/\partial{t})$, and employ the tunnelling
Hamiltonian as the perturbation. When $U_1=n_1\omega$ and
$U_2=n_2\omega$, the eigenvalue of $H_I$ are $0$. These represent
the zeroth-order approximations to the quasi-energies in the
perturbational expansion. The perturbation Hamiltonian $H_t$
induces the Floquet state transition. When the initial
localization is $|1\rangle$, the nonzero transition matrix element
is $P_{1,5^{'}}=\langle\langle{1}|H_t|5^{'}\rangle\rangle=
\sqrt{2}WJ_{n_1-n_2}(V/\omega)$. So the amplitude to frequency of
driving field is the roots of Bessel function of $n_1-n_2$ order,
the transition of initial localization state $|1\rangle$ is
strongly depressed. By using progressive formula of Bessel
function:
$J_m(x)\approx\sqrt{\frac{2}{\pi{x}}}\cos(x-\frac{\pi}{2}m-\frac{\pi}{4})$,
we find the roots of $J_{n_1-n_2}$ are also the roos of
$\sqrt{2J_{n_2}^2+6J_{n_1-n_2}^2}$ when $n_1-n_2$ is even.
Therefore, at the point of quasi-energies collapse, the
two-electron dynamical localization is set up.

We show above the phenomena of dynamical localization occurs at
the quasi-energy collapse when the two electrons are localized in
left QD. This quantum phenomena can also happen when the two
electrons are localized initially in centra QD. We verify this
assertion by calculating the quasi-energy and $BP_{min}$ in Figure
\ref{epsilon_LPmin_V}(d) with the parameters $U_1=20$, $U_2=4$,
$W=1$, $\omega=2$, the initial condition $C_2(0)=1$,
$C_\alpha(0)=0$ when $\alpha\neq2$. It shows that the features of
the Figure \ref{epsilon_LPmin_V}(d) are similar to that of the
Figure \ref{epsilon_LPmin_V}(b) for the strong field regime. The
calculation indicates that the two electrons of initial localized
in the centra QD can also maintain its localization at the
quasi-energy collapse. For the weak field regime, the effects of
Coulomb interaction emerge obviously. With the increase of $U_2$,
a series of quasi-energy crossing, and the dynamical localization
is set up at the quasi-energy crossing.

\section{SUMMARY}

In summary, applying three-site Hubbard model and Floquet theorem,
we have investigated the dynamic behaviors of two electrons
confined in a line-shape three quantum dot molecule driven by an
ac-field. We find the triplet state and triplet state sub-space
are decoupled fully, and can be investigated respectively. In
special, the triplet state subspace can be divided into 3
three-dimensional triplet state sub-space which are also
decoupled. The analysis indicates the dynamical properties in the
3 three-dimensional triplet state sub-space, as well as the
singlet state sub-space of no double-occupancy, are similar to
that driven two electrons in two QD molecule. The calculations and
analysis indicate that the dynamical properties in every subspace
can be divided into two regime: In the strong-field regime, the
dynamical behaviors are similar to that of non-interacting
electrons in superlattice, and in the weak-field regime, the
Coulomb interaction enhance the localization effect. We believe
the method and the results presented here are useful to explore
the effects of ac-field to control of multi-electron and multi-QD
system.

\begin{acknowledgements}
This work is supported by  by the Doctoral Fund of the Ministry of
Education of China Grand No. 96002703.
\end{acknowledgements}

\begin{table}[htbp]\caption{The two electron distribution in three QD molecule}
\tabcolsep0.03in
\begin{center}
\begin{tabular}{|c|c|c|c|c|c|c|c|c|c|c|c|c|c|c|c|}
\hline
&$|1\rangle$&$|2\rangle$&$|3\rangle$&$|4\rangle$&$|5\rangle$&
$|6\rangle$&$|7\rangle$&$|8\rangle$
&$|9\rangle$&$|10\rangle$&$|11\rangle$&$|12\rangle$&$|13\rangle$&$|14\rangle$&$|15\rangle$\\
\hline A&$\uparrow\downarrow$&&&$\uparrow$&$\uparrow$&$
\downarrow$&$\downarrow$&&&&&$\uparrow$&$\uparrow$&$
\downarrow$&$\downarrow$\\
\hline B&&$\uparrow\downarrow$&&$\uparrow$&$\downarrow
$&$\uparrow$&$\downarrow$&$\uparrow$&$\uparrow$&$\downarrow$&$\downarrow$&&&&\\
\hline C&&&$\uparrow\downarrow$&&&&&$\uparrow$&$
\downarrow$&$\uparrow$&$\downarrow$&$\uparrow$&$\downarrow$&$\uparrow$&$\downarrow$\\
\hline
\end{tabular}
\end{center}\label{two electron distribution}
\end{table}

\end{document}